\documentclass[twocolumn]{aastex701}

\newcommand{\xmm}{{\it XMM-Newton}}
\newcommand{\kepler}{{\it Kepler}}

\begin{document}

\title{Hunt for the mHz variability in the TESS and \xmm\ observations of nova-like cataclysmic variables}

\author{Andrej Dobrotka}
\affiliation{Advanced Technologies Research Institute, Faculty of Materials Science and Technology in Trnava, Slovak University of Technology in Bratislava, Bottova 25, 917 24 Trnava, Slovakia}
\email[show]{andrej.dobrotka@stuba.sk}

\author{Jozef Magdolen}
\affiliation{Advanced Technologies Research Institute, Faculty of Materials Science and Technology in Trnava, Slovak University of Technology in Bratislava, Bottova 25, 917 24 Trnava, Slovakia}
\email{jozef.madolen@stuba.sk}

\author{Martin Melicher\v{c}\'ik}
\affiliation{Advanced Technologies Research Institute, Faculty of Materials Science and Technology in Trnava, Slovak University of Technology in Bratislava, Bottova 25, 917 24 Trnava, Slovakia}
\email{martin.melichercik@stuba.sk}


\begin{abstract}

We analysed the flickering of selected nova-like cataclysmic variables observed by the TESS satellite and \xmm. We searched for break frequencies ($f_{\rm b}$) in the corresponding power density spectra (PDS), and for any long-term evolution. We found a new optical $f_{\rm b}$ in three nova-like systems and confirmed that the value of this frequency is clustered around 1\,mHz. V504\,Cen and V751\,Cyg show possible X-ray counterparts of $f_{\rm b}$ that had previously only been seen in MV\,Lyr. This points towards the very central disc for source localisation. We investigated a previously proposed correlation between white dwarf mass and $f_{\rm b}$, but thanks to the new measurements we do not conclude its existence. V3885\,Sgr and V1193\,Ori show flaring activity in the long-term light curve during which TESS observations were made. The corresponding PDSs show changes in shape and disappearance of $f_{\rm b}$. TT\,Ari and SGRt\,062340.2-265715 exhibit smooth changes in the long-term optical light curve, and the corresponding TESS observations show variable $f_{\rm b}$ during these changes. $f_{\rm b}$ is higher for lower brightness, which was seen only in MV\,Lyr so far.

\end{abstract}



\section{Introduction}
\label{introduction}

Nova-like systems are a subgroup of cataclysmic variables (CVs). These are interacting binaries powered by an accretion process. If the magnetic field of the central object is not strong enough, an accretion disc is present. The mass is transferred from a main sequence companion star via Roche lobe overflow, and flows towards the central white dwarf (WD) (see e.g. \citealt{warner1995} for a review).

CVs are either in high or low optical brightness states. These states depend on the mass accretion rate $\dot{m}_{\rm acc}$ through the disc. In the high state, $\dot{m}_{\rm acc}$ is high, the disc is bright and hot, and the hydrogen is fully ionised. When $\dot{m}_{\rm acc}$ decreases, the disc transitions to the low state, becoming cold, and the hydrogen recombines. The alternation between these two states is typical for dwarf novae and is driven by viscous-thermal instability (\citealt{osaki1974,hoshi1979,meyer1981}). In the high state, the accretion disc is (almost) fully developed up to the WD, while it is truncated in the low state. This truncation explains the delays between optical, UV radiation, and X-rays (see e.g. \citealt{schreiber2003}).

Nova-likes are CVs with a high mass transfer from the secondary star. The $\dot{m}_{\rm acc}$ is sufficient to maintain the accretion disc in a quasi-permanent high state. This is because $\dot{m}_{\rm acc}$ is above a critical value, and the viscous-thermal instability does not appear. However, some fluctuations in mass transfer from the secondary, probably due to star spots, occasionally appear, and the nova-likes experience a sudden transition to a low state for a relatively short time (\citealt{honeycutt2004}).


A typical manifestation of the underlying accretion process is fast stochastic variability, called flickering. The power density spectra (PDS) of this variability exhibit the shape of red noise or band limited noise, with characteristic frequencies in the form of a break $f_{\rm b}$ or Lorentzian (\citealt{scaringi2012a,dobrotka2016}). These characteristic frequencies provide information on the timescales of the physical processes that generate the variability. Thanks to new instruments such as the \kepler\ spacecraft, we know that the PDSs of CVs show a multi-component character. Well-studied cases are MV\,Lyr (\citealt{scaringi2012a}), V1504\,Cyg (\citealt{dobrotka2015}) and V344\,Lyr (\citealt{dobrotka2016}). X-ray observations have also revealed multicomponent PDSs in systems such as SS\,Cyg (\citealt{balman2012}), RU\,Peg (\citealt{dobrotka2014}) or MV\,Lyr (\citealt{dobrotka2017}).

\kepler\ observations are ideal for the search for $f_{\rm b}$. Thanks to the high quality data, it allows to track the evolution of this PDS parameter. \citet{dobrotka2020} measured the time evolution of $f_{\rm b}$ during the transition of MV\,Lyr from high to low optical state and back. The authors concluded that $f_{\rm b}$ is rising towards the low state, and it returns to its original value after the system returns to the high optical state. This tells a lot about the behavior of the accretion disc, but unfortunately it is hard to say whether this behavior is typical for nova-likes in general, because there are no other measurements of this phenomenon.

\citet{scaringi2014} proposed a sandwich model to explain the presence of the log($f$/Hz) $\simeq$ -3 signal detected in MV\,Lyr. This model consists of a standard geometrically thin and optically thick disc surrounded by a geometrically thick and optically thin disc in the central disc region. This inner thick disc or corona is hot and radiates in X-rays. Such a corona consists of evaporated gas from the underlying geometrically thin disc (\citealt{meyer1994}). The radiated X-rays are reprocessed into optical by the geometrically thin disc. The physical origin of the variability is explained by propagating mass accretion fluctuations (\citealt{lyubarskii1997,kotov2001,arevalo2006}) occurring within the corona.

The X-ray origin of the variability was confirmed by \xmm\ observations of the nova-like MV\,Lyr (\citealt{dobrotka2017}), where the dominant frequency close to log($f$/Hz) $\simeq$ -3 was also present in the X-rays. So far this is the only detection of a characteristic PDS frequency in optical and X-rays in a nova-like, showing consistent values and pointing towards a common physical origin. It supports the sandwich model interpretation, but it opens a question of whether such a correlation is present in nova-likes in general or if it is unique for MV\,Lyr.

Even if the sandwich model is known from X-ray binaries or AGNs (\citealt{dove1997}), it has a serious energetic problem in CVs. The ratio of X-ray to optical luminosity in these binaries is of the order of 0.1 (see e.g. \citealt{dobrotka2020}) or less (see e.g. \citealt{balman2014}). Clearly, this is too low to explain the observed optical variability with the reprocessing scenario. \citet{dobrotka2019} studied the flare profile of the flickering in the \kepler\ data of MV\,Lyr, and found two components with different amplitudes. Both components, the central spike and the side-lobes, have similar characteristic frequencies of log($f$/Hz) $\simeq$ -3, blending into the standard PDS and becoming indistinguishable. While the central spike is problematic to explain by the reprocessing due to too high amplitude, the low amplitude side-lobes match the scenario (\citealt{dobrotka2020}). This suggests that the variability may originate in two separate regions.

\citet{dobrotka2020} summarized various detections of $f_{\rm b}$ in the PDSs of CVs. The study implies that two characteristic frequencies can exist in the low state, while another frequency close to log($f$/Hz) $\simeq$ -3 appears in the high state. This hunt for $f_{\rm b}$ continued in \citet{dobrotka2024}. 15 new $f_{\rm b}$ were detected among nova-like systems in a high optical state with apparent concentration of values between log($f$/Hz) = -2.95 and -2.84. The probability that this signal close to 1\,mHz is not a random feature of otherwise uniform distribution is at least 96\%. Moreover, these $f_{\rm b}$ were detected in systems with low inclination. All PDSs with red noise without any $f_{\rm b}$ have considerably higher inclination. Finally, the authors also searched for correlation between the WD mass and $f_{\rm b}$, but the low number of measurements did not allow a definite conclusion.

Apparently, the origin of the flickering variability in both the optical and X-rays is not well understood yet. The number of measured $f_{\rm b}$ in PDSs is still low, or even extremely low in X-rays. In this work, we focus on the search for $f_{\rm b}$ in PDSs calculated from TESS and \xmm\ data of selected nova-like CVs. We focus on three different tasks; 1) to expand the number of measured $f_{\rm b}$ by \citet{dobrotka2024}, 2) to search for $f_{\rm b}$ correlation with optical state or its long-term behaviour, and 3) to search for $f_{\rm b}$ in X-rays.

In Section~\ref{section_observations} we describe the selection of systems and the reduction of observations both from TESS and \xmm. Sections~\ref{section_pds_tess} and \ref{section_pds_xmm} describe the PDS analysis of TESS and \xmm\ data and their results, respectively. These results are discussed and summarized in Sections~\ref{section_discussion} and \ref{section_summary}, respectively.

\section{Selected systems and observations}
\label{section_observations}

We first focused on TESS data of systems present in the Ritter's catalogue (\citealt{ritter2003}) but not analysed in \citet{dobrotka2024}. The latter analysed nova-like systems studied in \citet{bruch2022}, \citet{bruch2023a} and \citet{bruch2023b}, where superhumps were detected. Superhumps together with the orbital period contaminates the PDS, therefore the known presence of these signals allowed us to define a suitable frequency interval of the PDS to be studied. In this work, we selected SW\,Sex, UX\,UMa and VY\,Scl systems as subclasses of nova-likes (see \citealt{warner1995} for review) in order to avoid undesired objects classified as nova-likes. We avoid, for example, intermediate polars with truncated discs and strong spin frequency, which can affect the PDSs (see Fig.~2 in \citealt{dobrotka2024}). Intermediate polars and systems without disc, such as polars or compact binaries, such as AM\,CVn are also classified as nova-likes.

As a second step, we searched for \xmm\ observations of all objects analysed in \citet{dobrotka2024} and in this work.

As a third step, we selected \xmm\ observations of objects with target type "CV" and "nova-like" in the \xmm\ archive. However, these also include systems like magnetic polars without accretion disc, intermediate polars, AM\,CVn compact binaries with a hydrogen poor companion, or very old CVs known as period bouncers. We excluded them like in the TESS case and we focused on "standard" nova-like systems as studied in \citet{dobrotka2024}. We found observations of two systems not present in the TESS list from the previous step. SRGt\,062340.2-265715 was simply not listed in the previous catalogue output, and V442\,Oph does not have a TESS light curve.

For TESS light curves the Pre-search Data Conditioning Simple Aperture Photometry (PDCSAP) light curves were downloaded\footnote{Using Python Lightkurve library; \url{https://docs.lightkurve.org}.} from the Mikulski Archive for Space Telescopes (MAST) with a cadence of 120 seconds. The PDCSAP light curve is more suitable because systematic artifacts like long-term trends are removed.

For \xmm\ light curves the data were downloaded from the XMM-Newton Science Archive (XSA), and we obtained light curves with the Science Analysis Software (SAS), version 21.0. We focused on the EPIC/pn detector due to its higher throughput. The light curves were extracted from a circular region with a radius of 20\arcsec\ centred on the source, while the background was extracted from a region offset and the same radius. We used the {\tt epproc} tool to re-generate calibrated events files, and {\tt evselect} for light curve construction. We used a 50\,s time bin. Since timing analysis is limited by the duration of the observation even if the source is very bright and the S/N ratio is very good, we excluded all observations shorter than 10\,ks.

\section{PDS analysis of TESS data}
\label{section_pds_tess}


%
%

For PDS calculation, we followed the same procedure as in \citet{dobrotka2024}. We selected light curves without large gaps empirically defined as a duration of 0.2 days or more. Within these light curves with no large gaps, we selected light curve portions with a duration of $n_{\rm d}$ days. Due to relatively large PDS scatter, we empirically derived the optimal value of $n_{\rm d} = 10$. Not all of the examined objects have such uninterrupted 10 days intervals. Therefore, we redid the analysis also for $n_{\rm d} = 5$. We could use the whole light curve portion between the gaps, but we prefer to have all PDSs equivalent with the same frequency resolution and PDS bin step.

To estimate the PDS, we divided each light curve portion into $n_{\rm subs}$ subsamples. For each subsample, we calculated a periodogram (power\footnote{We used power normalized by fractional rms.} $p$ as a function of frequency $f$) using the Lomb-Scargle algorithm\footnote{We used python's package {\tt Astropy} (\citealt{astropy_collaboration2013,astropy_collaboration2018,astropy_collaboration2022}).} (\citealt{scargle1982}). The whole $f$ interval was re-binned with a constant frequency step of 0.1\,dex, and all log($p$) points within each frequency bin were averaged. Averaging of log($p$), rather than $p$, is recommended by \citet{papadakis1993}. A minimum number of averaged log($p$) values per bin equal to $n_{\rm min} \times n_{\rm subs}$ ($n_{\rm min}$ points from each of $n_{\rm subs}$ periodograms per bin) must also be fulfilled. If the condition is not met, the bin should be broadened until the condition is fulfilled. This ensures enough points for calculating the mean value with the standard error of the mean.

The lowest PDS frequency (before re-binning) we took proportional to the duration of the light curve subsamples (1/$n_{\rm d}$). Therefore, as $n_{\rm subs}$ and $n_{\rm min} \times n_{\rm subs}$ values, we chose 10 and $3 \times 10$, respectively. The high-frequency end of the periodogram is set by the Nyquist frequency.

\citet{dobrotka2024} focused on nova-like systems with known superhump periods, which are periods close to the orbital period. PDSs of such systems are contaminated by these signals, therefore the $f_{\rm b}$ search was limited to the interval from log($f$/Hz) = -3.5 to higher frequencies. Even though the systems selected in this work do not show superhumps, we kept the same studied frequency interval. It appears that PDSs of eclipsing systems are strongly contaminated by higher harmonics of the orbital period below log($f$/Hz) = -3.5.

Finally, we fitted the resulting PDSs with a broken power law model using the {\small GNUPLOT}\footnote{\url{http://www.gnuplot.info/}} software.

\subsection{Selection of positive detections}

First, we selected only successful and credible fits. First criterion was that a broken power law fit must converge to $f_{\rm b}$ between log($f$/Hz) = -3.5 and -2.4. If the PDS has a shape of a red noise, the broken power law is not able to do so, and $f_{\rm b}$ reached the lower end of the PDS. Second criterion was to have credible fits with acceptable uncertainty. We empirically selected 0.1 as an upper limit for $f_{\rm b}$ error. Larger errors resulted from too scattered PDS, and such $f_{\rm b}$ value is not credible. Therefore, all $f_{\rm b}$ with higher uncertainty than 0.1 were excluded.

After fitting all PDSs, and selecting successful and credible fits, we selected systems with positive detection of $f_{\rm b}$. To ensure that the $f_{\rm b}$ is real, and is not just a random fluctuation of a scattered PDS, we consider a positive detection of $f_{\rm b}$ in a system when at least 50\% ($n_{\rm p} = 0.5$) of all PDSs yield a successful fit. This limit is based on the well-studied system MV\,Lyr, where we are certain from the \kepler\ data that $f_{\rm b}$ is present. TESS PDSs of the same object yield a successful fit with $n_{\rm p} = 0.70$ and 0.77 for $n_{\rm d} = 5$ and $n_{\rm d} = 10$ (\citealt{dobrotka2024}), respectively. Ideally, we would rely on the same $n_{\rm p}$ values, but MV\,Lyr is a much brighter system compared to many other selected nova-likes, which makes detection easier. Therefore, a limit of $n_{\rm p} = 0.5$ is a good compromise.

Finally, having only two light curve portions and $n_{\rm p} = 0.5$ means that we have only one successful fit. This is not enough for a confident detection, therefore a minimal number of successful fits $n_{\rm m}$ we chose two.

Table~\ref{table_objects_tess} lists all selected nova-likes observed by TESS. Clearly, almost all systems with no detection have $n_{\rm p}$ parameter 0 or very close to 0. Only two objects have values of 0.13 and 0.24 implying that few PDSs have successful fits. Nevertheless, this is too low and far from $n_{\rm p} = 0.5$. Therefore, non-detection is practically unambiguous, and we found three systems with positive $f_{\rm b}$ detection.
\begin{table}
\caption{List of studied systems.}
\begin{center}
\begin{tabular}{lccc}
\hline
\hline
object & $n_{\rm p}$ & $n_{\rm p}$ & class\\
& $n_{\rm d} = 10$ & $n_{\rm d} = 5$ &\\
\hline
AH\,Pic			& wn & wn & UX\\
BP\,Lyn			& wn & wn & UX\\
CH\,CrB			& -- & low,wn & UX\\
CM\,Del			& 0 & 0 & UX\\
HL\,Aqr			& -- & 0 & UX\\
HQ\,Mon			& 0 & 0 & UX\\
HS\,0139+0559	& wn & wn & UX\\
HS\,0220+0603	& wn & wn & UX\\
HS\,0229+8016	& wn & wn & UX\\
HS\,0455+8315	& 0 & 0 & UX\\
HS\,0642+5049	& 0 & 0.13 & UX\\
HS\,1813+6122	& 0.24 & 0.04 & UX\\
IM\,Eri			& wn & wn & UX\\
IX\,Vel			& 0 & 0.04 & UX\\
OZ\,Dra			& -- & wn & UX\\
RW\,Sex			& -- & 0 & UX\\
SW\,Sex			& 0 & 0 & UX\\
TT\,Tri			& -- & 0 & UX\\
V345\,Pav		& 0 & wn & UX\\
V363\,Aur		& 0 & 0 & UX\\
V1315\,Aql		& low & 0 & UX\\
V1776\,Cyg		& wn & wn & UX\\
V3885\,Sgr		& low & 0.5 & UX\\
WX\,Ari			& 0 & 0 & UX\\
SRGt\,062340.2-265715			& no & 0.8 & UX\\
BO\,Cet			& wn & wn & SW\\
EV\,Lyn			& -- & wn & SW\\
Leo\,5			& wn & wn & SW\\
V347\,Pup		& 0 & 0 & SW\\
V482\,Cam		& 0 & 0 & SW\\
GS\,Pav			& -- & wn & VY\\
LQ\,Peg			& -- & 0 & VY\\
TW\,Pic			& 0 & 0.02 & VY\\
V380\,Oph		& -- & low & VY\\
V794\,Aql		& -- & low & VY\\
V1294\,Tau		& wn & wn & VY\\
VY\,Scl			& 1 & 0.5 & VY\\
VZ\,Scl			& -- & 0 & VY\\
\hline
\end{tabular}
\tablecomments{The systems are listed by their class with UX, SW and VY meaning UX\,UMa, SW\,Sex and VY\,Scl, respectively. "low" means "low number" because the total number of light curve portions is one, and this is not able to fulfill the condition $n_{\rm m} = 2$. "wn" means "white noise", and the corresponding PDS shows pure white noise or substabtial part of the PDS including log($f$/Hz) = -3 is dominated by white noise (flat PDS without any slope). $n_{\rm p}$ parameter naturally equals zero.}
\end{center}
\label{table_objects_tess}
\end{table}

\subsection{Results}

\begin{figure*}
\resizebox{\hsize}{!}{\includegraphics[angle=-90]{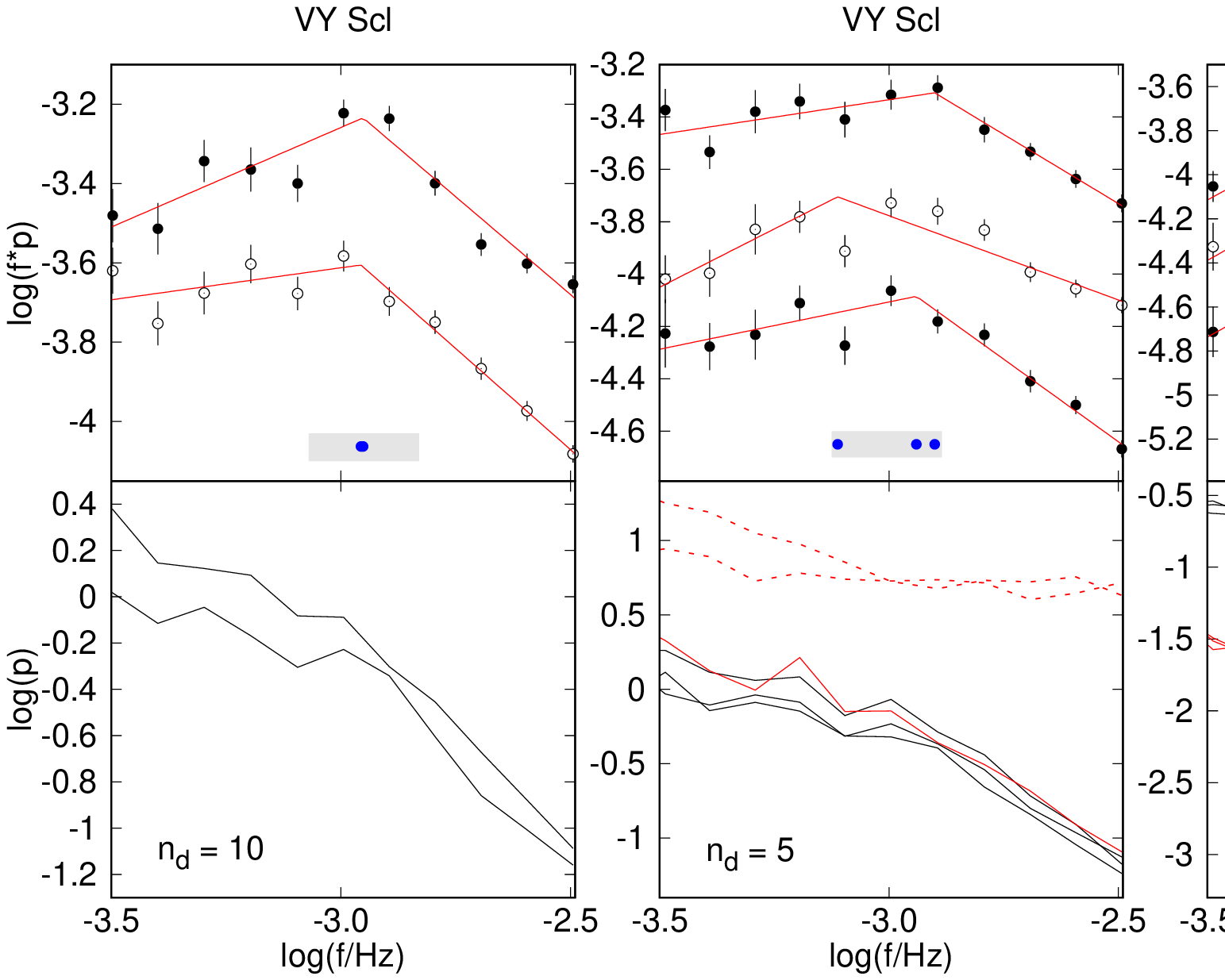}}
\caption{PDSs of three systems with positive detection of $f_{\rm b}$. Upper panel - PDSs with positive detection of $f_{\rm b}$ for $n_{\rm d} = 10$ and 5 light curves (points with errors of the mean). Red lines represent fits yielding $f_{\rm b}$ values with error lower than 0.1. PDSs are vertically offset for better visualisation. Individual $f_{\rm b}$ are shown as blue dots. The grey area around blue dots represents the natural scatter due to TESS light curve quality. The point style is varying just for clarity of the figure. Lower panel - all PDSs with (black lines) and without (red lines) positive detection of $f_{\rm b}$. Dashed lines represent low(er) optical state.}
\label{pds_tess}
\end{figure*}

Applying $n_{\rm m} = 2$ criterion, we got three positive detections; VY\,Scl, V3885\,Sgr and SRGt\,062340.2-265715. Case with $n_{\rm d} = 10$ is only VY\,Scl. All corresponding PDSs with broken power law fits are depicted in Fig.~\ref{pds_tess}. Individual $f_{\rm b}$ are shown as blue dots. The grey shaded area around these dots is the natural scatter of $f_{\rm b}$ due to noisy PDSs. This scatter is determined using the well-known MV\,Lyr case, and it represents the scatter of $f_{\rm b}$ calculated from TESS data (\citealt{dobrotka2024}). Clearly, all blue points except the SRGt\,062340.2-265715 case are within the scatter interval, therefore we do not expect any significant $f_{\rm b}$ variability between individual PDSs except SRGt\,062340.2-265715.

All measured $f_{\rm b}$ values with starting times of the light curve portions are listed in Table~\ref{table_lcs_tess}. We compare also $\chi^2_{\rm red}$ for a broken power law fit yielding $f_{\rm b}$ with simple red noise. Clearly, the red noise fits are unambiguously excluded. Seven broken power law fits yield $\chi^2_{\rm red}$ close to 1, while five are close to or higher than 1.5. The latter cases are natural for the limited performance of the TESS satellite and relatively scattered PDSs. Nevertheless, the broken power law fits are unambiguously better than simple red noise fits.

The weighted averages of $f_{\rm b}$ are summarized in Table~\ref{table_pds_prameters_systems}. Since SRGt\,062340.2-265715 has too scattered $f_{\rm b}$ measurements that imply variability, it must be taken with caution. The fitted log($f_{\rm b}$/Hz) values are $-3.23 \pm 0.02$, $-3.17 \pm 0.03$, $-2.94 \pm 0.03$ and $-3.08 \pm 0.03$. The first two values are consistent with the value of -3.2, while the last two $f_{\rm b}$ are higher. If we take also another PDS not satisfying conditions for a successful fit\footnote{Because of error of 0.13 which is slightly larger than the pre-define limit of 0.10.} it has a value of $-3.19 \pm 0.13$ also consistent with -3.2. Therefore, majority of fitted PDSs point toward log$(f_{\rm b}/{\rm Hz}) = -3.2$, and we use the first two successful fits for the weighted average. The two measurements with higher $f_{\rm b}$ will be discussed later.
\begin{table*}
\caption{Selected $f_{\rm b}$ for systems with positive detection using $n_{\rm d} = 10$ (first raw, above the line), and $n_{\rm d} = 5$ (below the line).}
\begin{center}
\begin{tabular}{lcccccc}
\hline
\hline
object & start & log($f_{\rm b}$/Hz) & $\chi^2_{\rm red}$ & start & log($f_{\rm b}$/Hz) & $\chi^2_{\rm red}$\\
& (MJD) & & BPL/RN & (MJD) & & BPL/RN\\
\hline
VY\,Scl			& 2088.245 & -2.952(046) & 2.57/13.79 & 2102.333 & -2.956(039) & 1.07/8.41\\
\hline
VY\,Scl			& 2088.245 & -2.900(045) & 0.78/5.59 & 2093.245 & -3.112(094) & 2.71/5.81\\
				& 2102.333 & -2.940(043) & 1.05/6.54 & & &\\
V3885\,Sgr		& 1658.373 & -3.062(034) & 1.71/17.74 & 1672.273 & -3.206(057) & 1.52/5.79\\
				& 1677.273 & -3.142(024) & 0.53/8.71 & & &\\
SRGt\,062340.2-265715	& 2206.737 & -3.226(017) & 0.48/14.71 & 2220.439 & -3.167(033) & 1.49/9.73\\
				& 3669.733 & -2.941(032) & 0.67/6.71 & 3683.695 & -3.080(030) & 0.74/6.11\\
\hline
\end{tabular}
\tablecomments{Start denotes the starting time of the selected light curve portion with MJD = JD - 2457000. The values in parenthesis represent the errors. For the $\chi^2_{\rm red}$ we quote values for broken power law (BPL) and simple red noise (RN).}
\end{center}
\label{table_lcs_tess}
\end{table*}
%
%
%

\begin{figure*}
\resizebox{\hsize}{!}{\includegraphics[angle=-90]{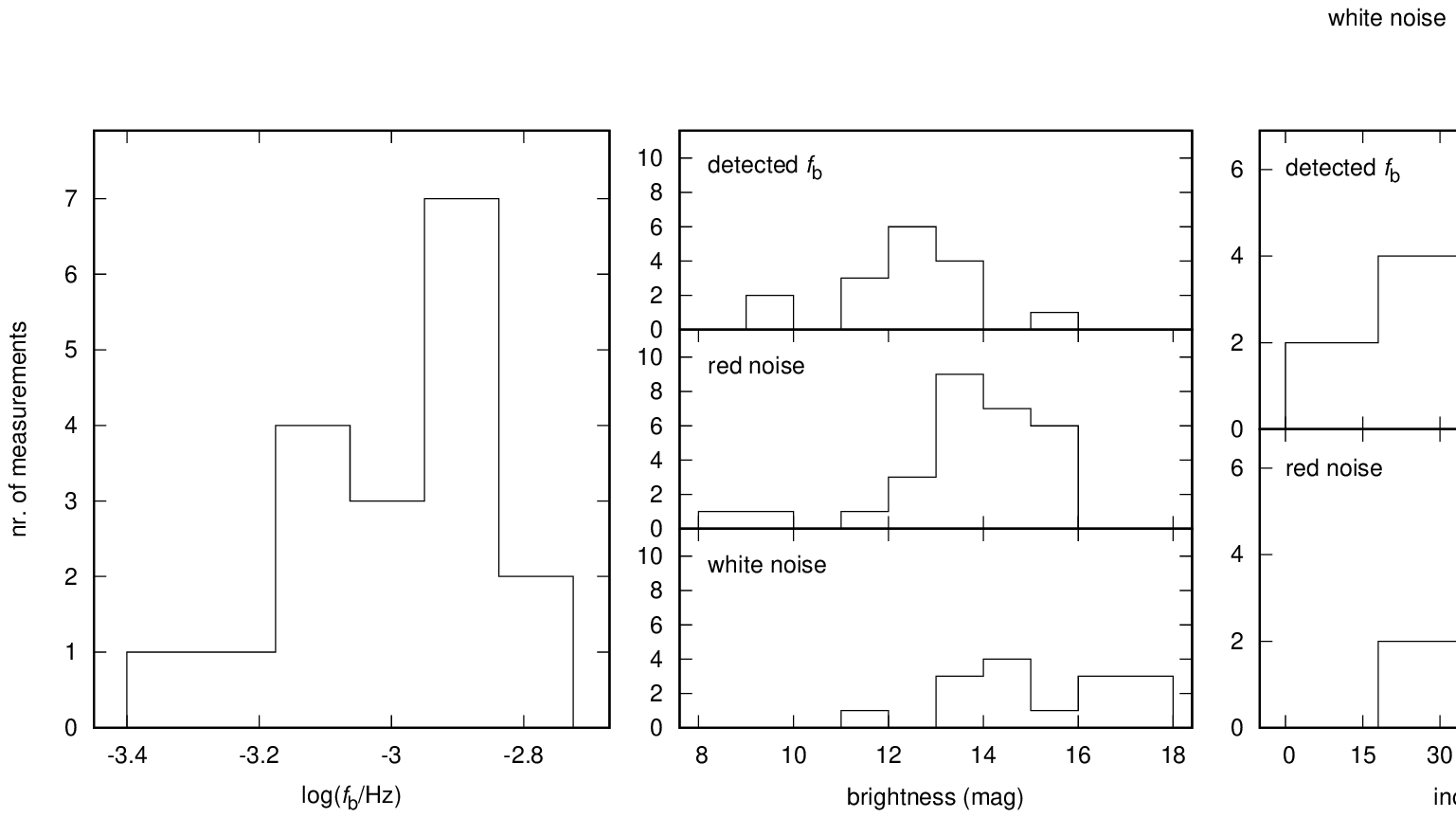}}
\caption{Number of nova-likes with detected and non-detected $f_{\rm b}$ for various parameters. The values are from this work and from \citet{dobrotka2024}. Left panel - number of detected $f_{\rm b}$ per frequency interval for nova-like CVs observed by TESS with condition $n_{\rm m} \geq 2$. Middle panel - Number of systems per magnitude interval with detected and non-detected $f_{\rm b}$. Right panel -  Number of systems brighter than 14\,mag per inclination interval with and without detection of $f_{\rm b}$.}
\label{hist}
\end{figure*}

Distribution of all detected $f_{\rm b}$ together with those from \citet{dobrotka2024} is depicted in the left panel of Fig.~\ref{hist}.
\begin{table*}
\caption{Weighted averages of measured $f_{\rm b}$ for individual systems with positive TESS detection (from this work and from \citealt{dobrotka2024} for direct comparison with \xmm\ measurements) and from \xmm\ observations.}
\begin{center}
\begin{tabular}{lccc}
\hline
\hline
object & log($f_{\rm b}$/Hz) & $n_{\rm d}$ & log($f_{\rm b}$/Hz)\\
& TESS & & \xmm\ \\
\hline
VY\,Scl					& $-2.954 \pm 0.030$ & 10 & --\\
						& $-2.940 \pm 0.030$ & 5 & --\\
V3885\,Sgr				& $-3.126 \pm 0.019$ & 5 & --\\
SRGt\,062340.2-265715	& $-3.213 \pm 0.015$ & 5 & --\\
\hline
V504\,Cen				& $-2.860 \pm 0.046$ & 10 & $-3.003 \pm 0.332$\\
V751\,Cyg				& $-2.982 \pm 0.023$ & 10 & $-2.991 \pm 0.134$\\
\hline
\end{tabular}
\tablecomments{Below the line are two TESS measurements from \citet{dobrotka2024}.}
\end{center}
\label{table_pds_prameters_systems}
\end{table*}
%
%
%
%

\section{PDS analysis of \xmm\ data}
\label{section_pds_xmm}


We used the same method for PDS estimate like for TESS data. The main difference is the duration of the observations. With \xmm\ we don't have the luxury of selecting 5 or 10 days portions, and to divide these light curves into 10 subsamples. Typical duration of \xmm\ observation is between 20 and 40\,ks (based on the histogram of selected observations). This is 0.23 to 0.46 day, which is incomparable with TESS portions. Apparently, for \xmm\ observation we must use different divisions. Moreover, for TESS case we have more such light curve portions, therefore we can define what is a positive detection. For \xmm\ we must usually rely on a single observation.

For \xmm\ PDS estimate we used $n_{\rm subs} = 2$ and 3, and as a minimal number of points per bin we used 6 ($n_{\rm subs} \times 2$ or 3). We searched for $f_{\rm b}$ using the same fit as for TESS data. Since orbital and superhump periods are not standardly seen in X-rays, we fitted the PDSs from log($f$/Hz) = -4.0 to -2.0.

\begin{figure}
\includegraphics[width=200mm,angle=-90]{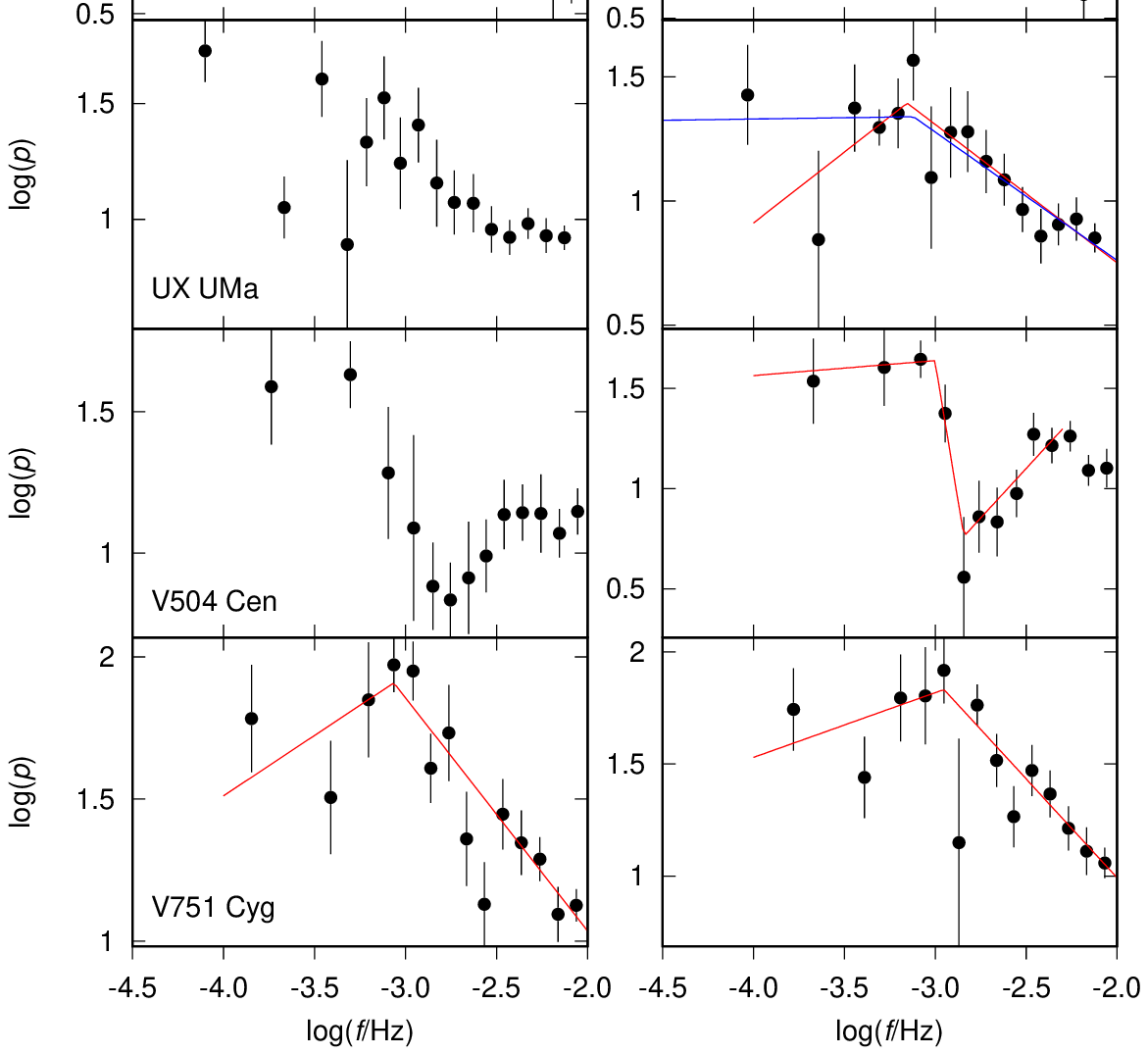}
\caption{PDSs calculated from \xmm\ observations using $n_{\rm subs} = 2$ (left column) and 3 (right column). Red lines are the fits to the PDSs with lower PDS end of log($f$/Hz) = -4.0 while blue is from log($f$/Hz) = -4.5.}
\label{pds_xmm}
\end{figure}

All PDSs calculated from \xmm\ data are shown in Fig.~\ref{pds_xmm}. All ObsIDs of analysed observations and measured $f_{\rm b}$ values are listed in Table~\ref{table_lcs_xmm}. We also quote the corresponding $\chi^2_{\rm red}$ for the broken power law and a simple red noise model. The weighted averages of the candidates $f_{\rm b}$ are summarized and compared to the TESS detections (if any) in Table~\ref{table_pds_prameters_systems}.
\begin{table*}
\caption{Analysed \xmm\ observations and measured $f_{\rm b}$.}
\begin{center}
\begin{tabular}{lcccccc}
\hline
\hline
object & ObsID & log($f_{\rm b}$/Hz) & $\chi^2_{\rm red}$ & log($f_{\rm b}$/Hz) & $\chi^2_{\rm red}$ & duration\\
& & $n_{\rm subs} = 2$ & BPL/RN & $n_{\rm subs} = 3$ & BPL/RN & (ks)\\
\hline
IX\,Vel					& 0111971001 & -- & -- & -- & -- & 19.0\\
TW\,Pic					& 0500970101 & -- & -- & -- & -- & 41.4\\
						& 0500970301 & -- & -- & -- & -- & 16.0\\
UU\,Aqr					& 0930800101 & -- & -- & -- & -- & 17.4\\
V442\,Oph				& 0305440801 & -- & -- & -- & -- & 11.5\\
SRGt\,062340.2-265715	& 0901220101 & $-2.984 \pm 0.118$ & 2.61/5.18 & $-3.008 \pm 0.140$ & 1.87/4.66 & 44.1\\
						&            & $-2.347 \pm 0.060$ & 2.09/4.83 & $-2.445 \pm 0.085$ & 2.05/4.33 &\\
V1084\,Her				& 0804110101 & $-3.051 \pm 0.191$ & 1.37/1.48 & $-2.976 \pm 0.300$ & 1.38/1.43 & 21.0\\
UX\,UMa					& 0084190201 & -- & -- & $-3.155 \pm 0.104$ & 0.57/0.80 & 45.9\\
						&            & -- & -- & $-3.120 \pm 0.203$ & 0.64/0.81 &\\
V504\,Cen				& 0311590401 & -- & -- & $-3.003 \pm 0.332$ & 0.75/3.76 & 19.9\\
V751\,Cyg				& 0679580201 & $-3.064 \pm 0.226$ & 1.57/2.54 & $-2.952 \pm 0.166$ & 0.96/2.12 & 25.6\\
\hline
\end{tabular}
\tablecomments{Duration is the duration of the EPIC/pn light curve. If two values are quoted for one ObsID, the first is for fitting interval from log($f$/Hz) = -4.0, and the second is from log($f$/Hz) = -4.5. For the $\chi^2_{\rm red}$ we quote values for broken power law (BPL) and simple red noise (RN).}
\end{center}
\label{table_lcs_xmm}
\end{table*}

The first four PDSs in Fig.~\ref{pds_xmm} do not show any structure indicating the presence of $f_{\rm b}$. TW\,Pic has two observations. Since both have the same characteristics, we show only the longer-one. All PDSs from the four nova-likes are consistent with a red noise shape, only the V442\,Oph case is too short (11.5\,ks) to confirm the pure red noise nature of the PDS. Otherwise, the non-detection of any $f_{\rm b}$ in these systems is consistent with conclusions from the TESS data.

Some deviations from simple red noise are possible for SRGt\,062340.2-265715. The $\chi^2_{\rm red}$ are much better for broken power laws, but the values are still too large. For consistency check we fitted the PDS also from log($f$/Hz) = -4.5, and the resulting $f_{\rm b}$ is different. Apparently, the PDS does not show any robust trend, and we conclude no detection for this object. Another PDS with possible $f_{\rm b}$ is seen in V1084\,Her. However, $\chi^2_{\rm red}$ are practically the same for the broken power law model and red noise. Therefore, we conclude no detection in this case too. UX\,UMa shows very scattered PDS with $n_{\rm subs} = 2$ but a changing red noise slope is seen for $n_{\rm subs} = 3$. Fit with lower PDS end of log($f$/Hz) = -4.5 yields consistent $f_{\rm b}$ value with the original fit. However, even if the $\chi^2_{\rm red}$ from the broken power law are lower than in the red noise case, the difference is small and lower than one. This suggests overfitting due to large PDS errors. No conclusion can be derived, and we conclude no detection. The non-detections of $f_{\rm b}$ in V 1084\,Her and UX\,UMa agree with the non-detections in TESS data.

The last two examples represent potential detections of $f_{\rm b}$. Deviations from a simple red noise are noticeable for V504\,Cen. The $n_{\rm subs} = 2$ case shows a possible change of the red noise trend at approximately log($f$/Hz) = -3.31, but the different slope towards the lower frequencies is represented by only one single point. This is insufficient for any fitting. The changing slope is better seen and represented by more PDS points in the $n_{\rm subs} = 3$ case. Due to the complicated PDS shape toward higher frequencies we used three linear functions and changed the higher frequency end for fitting to log($f$/Hz) = -2.2. Differences in $\chi^2_{\rm red}$ clearly support the broken power law model. The fitted $f_{\rm b}$ has relatively large error but the value of log($f_{\rm b}$/Hz) = $-3.00 \pm 0.33$ agrees with the TESS detection in \citet{dobrotka2024}. Finally, both $n_{\rm subs} = 2$ and 3 of V751\,Cyg show a clear change of the red noise slope close to log($f$/Hz) = -3. $\chi^2_{\rm red}$ supports broken power law as a better model mainly for $n_{\rm subs} = 3$. The averaged value of log($f_{\rm b}$/Hz) = $-2.99 \pm 0.13$ agrees with the TESS detection in \citet{dobrotka2024}.

\section{Discussion}
\label{section_discussion}


We searched for $f_{\rm b}$ of the flickering of nova-like CVs in optical and X-ray data from TESS and \xmm\, respectively. We found three new $f_{\rm b}$ in optical and two new $f_{\rm b}$ in X-rays. The latter agrees with previously detected optical values.

\subsection{PDS break frequency detection}
\label{section_break frequency detection}

\citet{dobrotka2020} found a possibility that CVs in a high state show a preferred characteristic PDS frequency close to log($f$/Hz) = -3. The authors summarized $f_{\rm b}$ measurements of five systems, with only three showing values close to log($f$/Hz) = -3. One is the nova-like MV\,Lyr, and the two remaining are dwarf novae V1504\,Cyg and V344\,Lyr. Apparently, the number of studied systems was low, therefore, it is still possible that the weak peak in the corresponding histogram is a random result of an otherwise uniform distribution. The authors simulated such histograms and found that the probability\footnote{Based on number of simulated histograms with the same or higher peaks as the observed histogram.} that the peak is not random is only 69\%.

The systematic search for $f_{\rm b}$ in selected nova-like CVs using TESS data performed by \citet{dobrotka2024} yields a much larger data sample than in \citet{dobrotka2020}. In addition to the known MV\,Lyr, the authors found 14 other new detections. The resulting distribution of $f_{\rm b}$ shows a concentration of values between log($f$/Hz) = -2.95 and -2.84 with a probability of 99\% not being a result of a random process. This probability is for the same criterion ($n_{\rm m} = 2$) as used in this work. The three new measurements from this work slightly increase the number of detected $f_{\rm b}$, but not in the potentially preferred interval between log($f$/Hz) = -2.95 and -2.84. The confidence of 99\% decreases to 96.6\%. Clearly, based on Fig.~\ref{hist} the $f_{\rm b}$ distribution is continuous and not structured into several values like discussed by \citet{dobrotka2020}. The potentially preferred value is probably just a maximum of the distribution.

$f_{\rm b}$ derived from \xmm\ observations have larger errors compared to the TESS results. This is due to a very short observation duration which is typical for \xmm. The main result of this analysis is that optical $f_{\rm b}$ have a possible X-ray counterpart in V504\,Cen and V751\,Cyg. So far this was demonstrated only for MV\,Lyr by \citet{dobrotka2017}.

\subsection{Break frequency vs long-term light curve}

If a system exhibits $f_{\rm b}$, not all PDSs yield its detection. This is due to the natural scatter of PDS points. Usually, such PDSs do not show any significant shape differences. However, \citet{dobrotka2020} showed that PDS exhibits a change in its morphology, and $f_{\rm b}$ can disappear during the transition from high to low optical state and vice versa. Therefore, non-detection of $f_{\rm b}$ can have a real physical reason. Instead of relying only on $f_{\rm b}$ detection or non-detection, we discuss also the shape of the PDS.

\begin{figure*}
\resizebox{\hsize}{!}{\includegraphics[angle=-90]{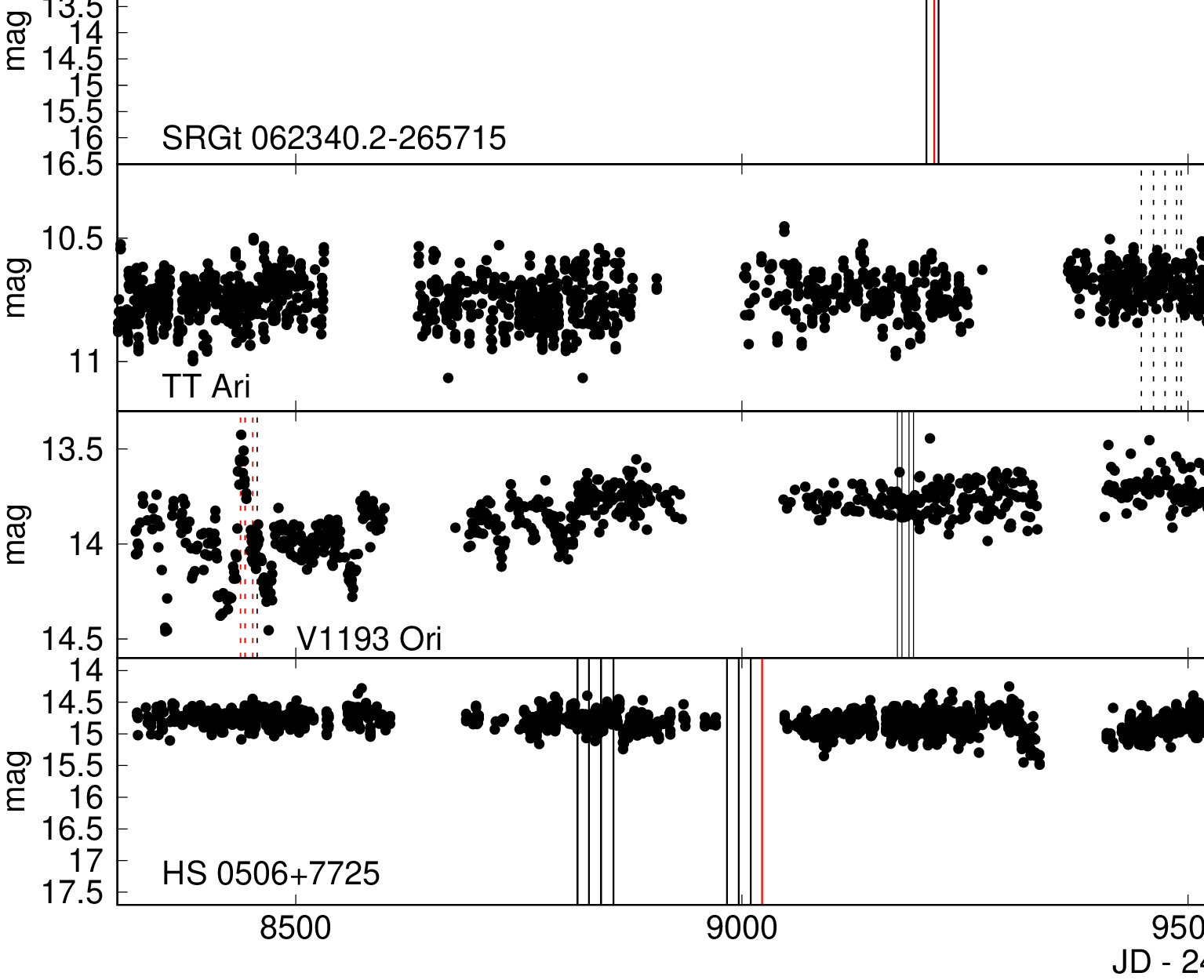}}
\caption{ASAS-SN light curves of selected systems with detected $f_{\rm b}$. Vertical lines represent time of TESS light curve portions with detected (black lines) and non-detected (red lines) $f_{\rm b}$. Blue vertical line shows the time of \xmm\ observation. Dashed lines represent observations taken during specific times discussed in the text (corresponding to dashed PDSs in Fig.~\ref{pds_tess} and \ref{pds_tt_v1193_hs}).}
\label{lc_asassn}
\end{figure*}

All three nova-likes with newly detected $f_{\rm b}$ show variability of PDS morphology. The shape and normalised power levels differ for PDSs with and without detection of $f_{\rm b}$ in VY\,Scl and V3885\,Sgr. In VY\,Scl two groups of PDSs are seen. The red dashed lines in Fig.~\ref{pds_tess} show a simple white noise implying absence of any flickering with typical red noise. As shown in the ASAS-SN light curve in Fig.~\ref{lc_asassn}, the corresponding observations were taken during the low state where the object is very faint and TESS is not able to detect any variability. Otherwise, this would be an excellent opportunity to study $f_{\rm b}$ difference between high and low optical state. One PDS without $f_{\rm b}$ detection is also taken during the high state (red solid line), but this is only because of error 0.11, which is slightly above the criterion we used for a successful fit.

V3885\,Sgr shows a different behaviour. The PDS power is considerably lower when PDS has a clear red noise shape without any $f_{\rm b}$ detection. Long-term ASAS-SN light curve in Fig.~\ref{lc_asassn} shows flaring activity and individual TESS observations are taken during different stages of this flaring. $f_{\rm b}$ is detected during a temporarily higher state seen as a local rebrightening at approximately JD - 2450000 = 8650. All other TESS observations outside of this rebrightening show simple red noise. This suggests a real change in flickering characteristics. Similar flaring variability has been observed in VY\,Scl systems when the disc is in the high state, for example in V794\,Aql (\citealt{honeycutt1994}, \citealt{honeycutt1998}), FY\,Per (\citealt{honeycutt2001}) or V4743\,Sgr (\citealt{dobrotka2021}). Either the flares are due to disc instabilities when mass transfer from the secondary temporarily ceases (\citealt{honeycutt1994}), or the disc remains in the high state while the variability is caused by mass transfer variations from the secondary (\citealt{honeycutt1998}). Varying mass transfer naturally generates changes in the accretion disc flow. This variable flow affects the structure of the disc, which is seen as changes or disappearance of $f_{\rm b}$.

Finally, as already mentioned, we see a drift of $f_{\rm b}$ for SRGt\,062340.2-265715. The $f_{\rm b}$ values are -3.23, -3.17, -2.94, and -3.08. The first observations with log($f_{\rm b}$/Hz) equal to -3.23, -3.17 (one additional red line at -3.19) are taken during a high state (upper panel of Fig.~\ref{lc_asassn}). A transition occurs approximately at JD - 2450000 = 10400, the system returns back to the high state, and another transition starts subsequently. Two TESS observations yielding log($f_{\rm b}$/Hz) of -2.94 and -3.08 are taken during this second transition (dashed lines in Fig.~\ref{lc_asassn}). \citet{dobrotka2020} showed that during the transition of MV\,Lyr to the low state the $f_{\rm b}$ rises. Therefore, SRGt\,062340.2-265715 is a second case where such rising $f_{\rm b}$ during a transition to the low state was observed.

\citet{dobrotka2024} did not study such PDS morphology, and focused only on $f_{\rm b}$ detection. The authors took into account the long-term light curve only if $f_{\rm b}$ showed an anomalous value. Therefore, we redid the PDS calculation of the objects the authors studied, and we searched for any systematic PDS shape change like found in this work. We found three systems with such behavior; TT\,Ari, V1193\,Ori and HS\,0506+7725.

\begin{figure}
\resizebox{\hsize}{!}{\includegraphics[angle=-90]{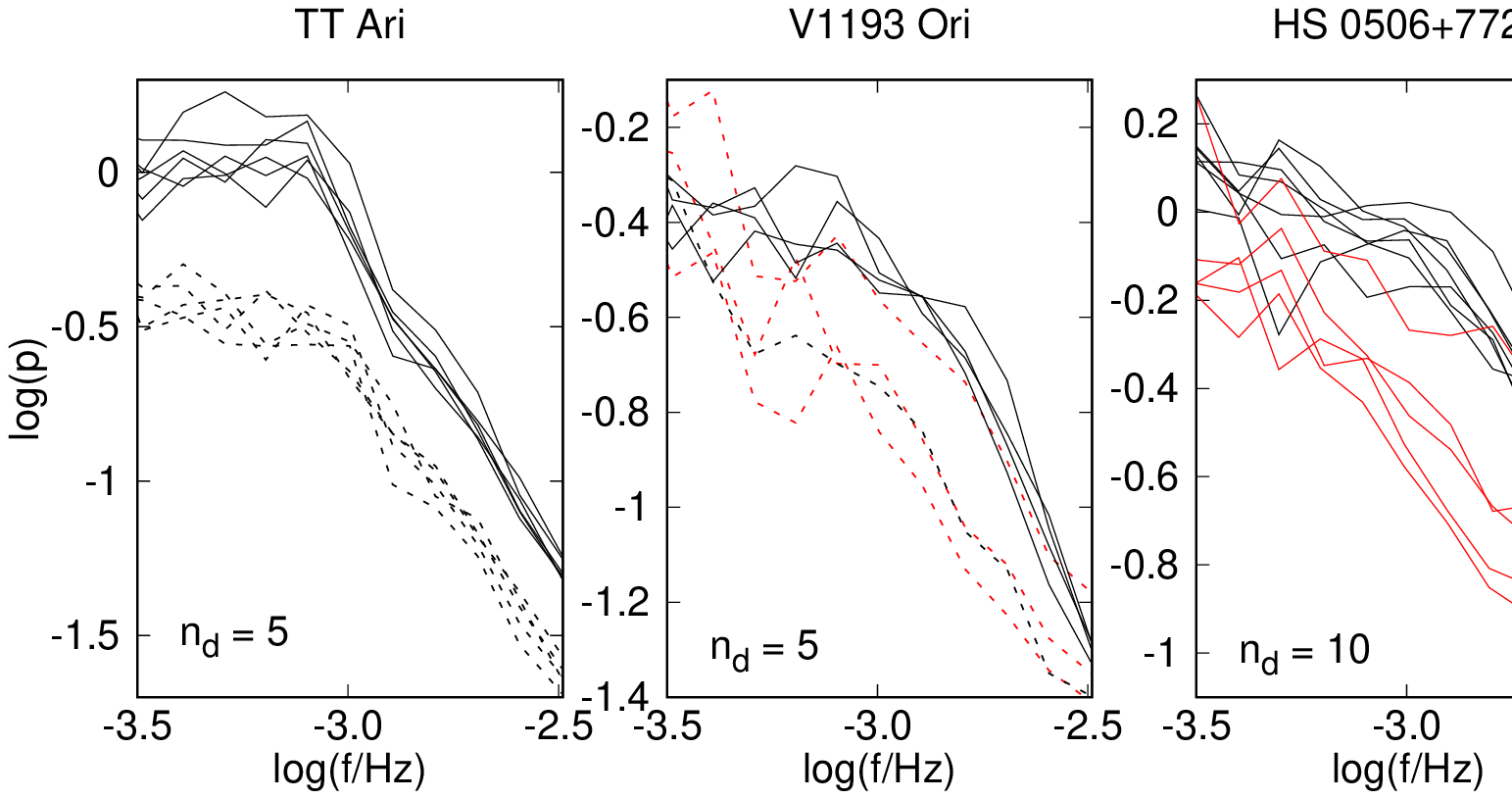}}
\caption{PDSs of TT\,Ari, V1193\,Ori and HS\,0506+7725. Black lines represent PDSs with $f_{\rm b}$ detection, and red lines without. Dashed lines represent PDSs taken during specific times shown as dashed lines in Fig.~\ref{lc_asassn} (see text for details).}
\label{pds_tt_v1193_hs}
\end{figure}

PDSs of TT\,Ari and its ASASN-SN light curve are shown in Figs.~\ref{pds_tt_v1193_hs} and \ref{lc_asassn}, respectively. The long-term light curve does not show any flaring like in V3885\,Sgr, nor the transition typical for VY\,Scl, but a noticeable difference in brightness level and amplitude of variability during the TESS observations is seen instead. Clear are two "groups" of PDSs. Lower PDS power is seen before JD - 2450000 = 9500. Subsequently, the brightness slightly decreases and after JD - 2450000 = 10000 it increases back or even to a slightly higher brightness level. TESS observations taken during this second stage show higher power suggesting a higher amplitude of variability. This higher amplitude is seen as considerably higher data scatter during the second group of TESS observations. Based on \citet{dobrotka2024} the brighter stage (later observations) yields lower values of $f_{\rm b}$. This behaviour is in agreement with the conclusion from \citet{dobrotka2020} where $f_{\rm b}$ showed lower values for a higher optical state. Apparently, the accretion exhibits some fluctuation after JD - 2450000 = 9500, and affects the PDS power with $f_{\rm b}$ value.

V1193\,Ori PDSs in Fig.~\ref{pds_tt_v1193_hs} shows three PDSs slightly offset towards lower power. This is clear mainly for higher frequencies where $f_{\rm b}$ is located. The PDS power at low frequencies is more or less similar. This is very similar to the PDS change during state transitions in MV\,Lyr (\citealt{dobrotka2020}). ASAS-SN light curve (Fig.~\ref{lc_asassn}) shows one flare during which the PDS did not yield $f_{\rm b}$ detection. Once the brightness level is stabilized, the PDSs yield detection of $f_{\rm b}$. The dashed lines in Fig.~\ref{pds_tt_v1193_hs} depict the PDSs taken during or close to the flare at approximately JD - 2450000 = 8400. No clear correlation between optical states is seen, but flaring activity indicates a sort of instability during which the accretion disc experiences changes. Then the PDS is affected and $f_{\rm b}$ is modified or even disappears.

The final and most challenging case is HS\,0506+7725. The PDSs in Fig.~\ref{pds_tt_v1193_hs} show a similar behaviour as in the V 1193\,Ori case. The $f_{\rm b}$ disappeared or decreased in power while the low frequency parts of all PDSs are more or less comparable. However, no state transition is seen in the long-term light curve (Fig.~\ref{lc_asassn}). Any statement about state transition is still not conclusive because eight observations were taken during gaps and we don't see how the brightness evolved. We cannot rule out any flaring activity during the four red PDSs seen in Fig.~\ref{pds_tt_v1193_hs}.

Finally, we conclude that variable optical brightness generates PDS change and $f_{\rm b}$ modification or even disappearance. Such variable brightness or flaring indicates fluctuations in mass transfer rate, and significant changes in the accretion flow are a consequence. Naturally, $f_{\rm b}$ as a finger-print of the accretion flow must answer. If $f_{\rm b}$ variability is detected, its value increases with decreasing brightness, as concluded by \citet{dobrotka2020}. Apparently, this behaviour is not typical only for MV\,Lyr but probably for more or all nova-likes. This suggests a common physical origin of the fast variability in these objects.

\subsection{Dependence on system parameters}

\begin{figure}
\resizebox{\hsize}{!}{\includegraphics[angle=-90]{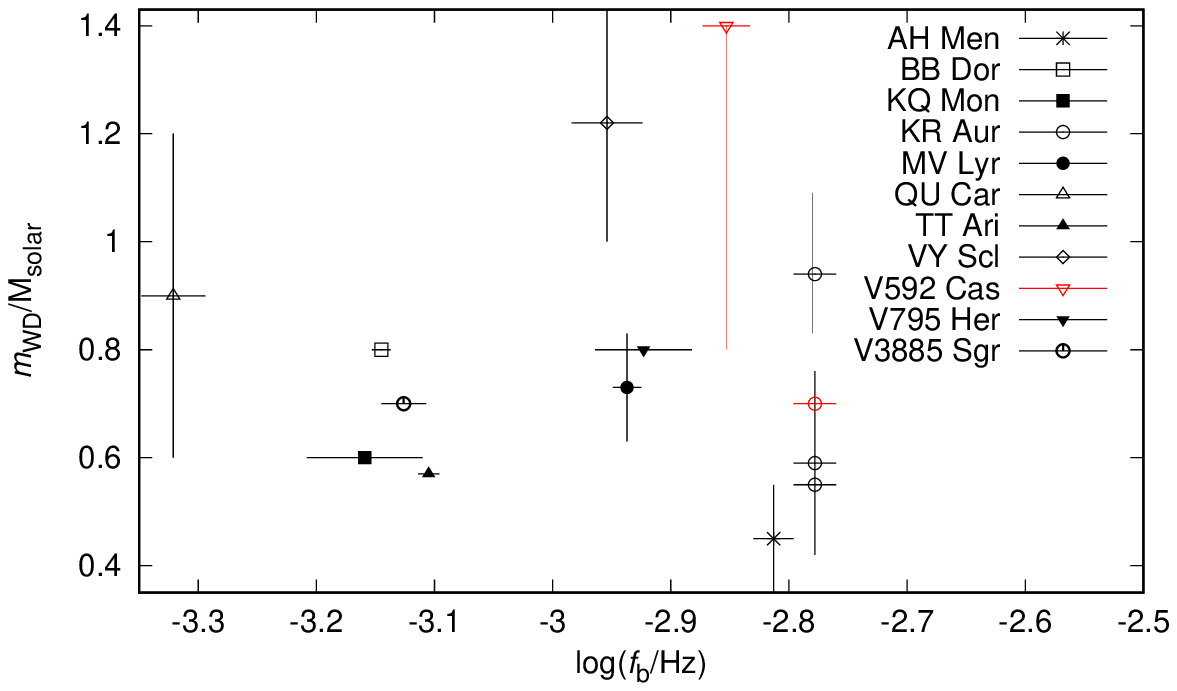}}
\caption{$m_{\rm WD}$ vs $f_{\rm b}$ from \citet{dobrotka2024} together with VY\,Scl and V3885\,Sgr studied in this paper. The red color represents uncertain measurements (see \citealt{dobrotka2024} for details).}
\label{wdmass_frekv}
\end{figure}

\citet{dobrotka2024} compared WD masses $m_{\rm WD}$ to detected $f_{\rm b}$ values. They proposed a possibility that these two parameters are correlated. However, the number of measured $f_{\rm b}$ was low, but mainly the scatter and uncertainties of $m_{\rm WD}$ are high. After adding VY\,Scl and V3885\,Sgr with $m_{\rm WD}$ from Table~\ref{table_system_prameters}, we see no such correlation (Fig.~\ref{wdmass_frekv}).
\begin{table}
\caption{System parameters of selected systems.}
\begin{center}
\begin{tabular}{lll}
\hline
\hline
object & incl. & $m_{\rm WD}$\\
& (deg) & (M$_{\rm \odot}$)\\
\hline
VY\,Scl & 30$^{(a)}$ & $1.22 \pm 0.22^{(a)}$\\
V3885\,Sgr & 45-75$^{(b)}$ & 0.70$^{(c)}$\\
\hline
CM\,Del & 73$^{(a)}$\\
HL\,Aqr & 19 - 27$^{(e)}$\\
IX\,Vel & 57$^{(a)}$\\
RW\,Sex & 34$^{(a)}$\\
SW\,Sex & ecl\\
TT\,Tri & ecl\\
TW\,Pic & $<$40$^{(f)}$\\
VZ\,Scl & ecl\\
WX\,Ari & ecl\\
V345\,Pav & ecl\\
V347\,Pup & ecl\\
V363\,Aur & ecl\\
V380\,Oph & 42$^{(a)}$\\
V482\,Cam & ecl\\
V794\,Aql & 60$^{(g)}$\\
V1315\,Aql & ecl\\
\hline
\end{tabular}
\tablecomments{First part is for systems with $f_{\rm b}$ detection, while second part is only for systems with red noise PDSs. Eclipsing (ecl) systems were identified in this work based on the morphology of the light curve.\\
$^{\rm (a)}$\citet{ritter2003}, $^{\rm (b)}$\citet{ribeiro2007}, $^{\rm (c)}$\citet{linnell2009}, $^{\rm (d)}$\citet{aungwerojwit2005}, $^{\rm (e)}$\citet{rodriguez2007}, $^{\rm (f)}$\citet{buckley1990}, $^{\rm (g)}$\citet{godon2007}}
\end{center}
\label{table_system_prameters}
\end{table}

Another possible correlation is detection of $f_{\rm b}$ and the inclination of the binary. \citet{dobrotka2024} found that systems with detected $f_{\rm b}$ have lower inclination than 60-75$^{\circ}$. Table~\ref{table_system_prameters} lists the inclinations of the systems studied in this work. Apparently, the systems with detected $f_{\rm b}$ confirm the conclusion of \citet{dobrotka2024} that low inclination is required for $f_{\rm b}$ detection. However, the second part of the table already discredits the original hypothesis that non-detection of $f_{\rm b}$ is due to high inclination, where the central disc as a source of the flickering is hidden by the disc edge. We found that five systems do have lower inclination than 60$^{\circ}$ and do not show any $f_{\rm b}$.

A possible explanation can be simply based on limiting capabilities of TESS. MV\,Lyr is a well-studied system with clear $f_{\rm b}$ and it does not show it in all TESS light curve portions (\citealt{dobrotka2024}). It is present only in 70\% or 77\% cases for $n_{\rm d} = 5$ and $n_{\rm d} = 10$, respectively. For example, TW\,Pic PDS shows a weak bend or hump close to log($f$/Hz) = -2.95. Fitting the PDSs from log($f$/Hz) = -3.35, we found five similar PDSs yielding consistent $f_{\rm b}$ between log($f$/Hz) = -2.9 and -3.1. Some have errors larger and some lower than 0.1. These features may be a weak manifestation of the presence of $f_{\rm b}$. Apparently, a stronger instrument and more detailed PDS is needed.



Therefore, it is possible that in faint systems $f_{\rm b}$ is simply not detected. High inclination CVs tend to be fainter which could explain why $f_{\rm b}$ is not detected for these inclinations. In the middle panel of Fig.~\ref{hist} we compare histograms of nova-likes per magnitude interval for three classes of objects\footnote{We used magnitudes from \citet{ritter2003} for objects from Table~\ref{table_objects_tess} of this work and Table~2 from \citet{dobrotka2024}.}. The brightness span for systems with detected $f_{\rm b}$ is from 9.5 to 15.1\,mag. Taking all systems showing only red noise the brightness span is from 9.0 to 15.7\,mag, therefore very similar. Contrary, taking systems dominated by white noise the brightness span is offset to higher values from 11.1 to 18.0\,mag. Apparently, there is no significant difference in brightness span for objects with $f_{\rm b}$ and red noise, but the bulk of distribution for red noise systems is offset toward lower magnitudes. This suggests that the brightness is not decisive, but probably plays an important role, and the non-detection of $f_{\rm b}$ is strongly biased by TESS capabilities.

Taking into account the brightness conclusion, we can set a limiting magnitude for histogram construction in the right panel of Fig.~\ref{hist}\footnote{If the inclination is uncertain in an interval we used the averaged value. If the inclination has an upper limit, we used the upper limit. For systems with detected $f_{\rm b}$, we did not apply the brightness filter.}. Taking 14\,mag as the limit, where the majority of the distribution of systems with detected $f_{\rm b}$ ends, there are HL\,Aqr and RW\,Sex with low inclination showing red noise. The case for red noise systems is not conclusive, but it appears that systems with detected $f_{\rm b}$ are not high inclination binaries.

\subsection{Physical origin of the break frequency}

Based on this work and \citet{dobrotka2017} we can conclude that three nova-likes show the same $f_{\rm b}$ in optical and X-rays. This suggests a common physical origin of the fast variability in these objects. The X-ray nature points toward the central corona or the boundary layer between the inner disc and the WD as the source of the variability. In the case of the corona, it supports the sandwich model proposed for MV\,Lyr by \citet{scaringi2014}. However, additional information is needed to discriminate between the two scenarios.

A boundary layer could be connected to the inner disc radii. During the transition to a low state, the mass accretion drops, the inner disc starts to be truncated and the inner radius of the optically thick disc increases like in dwarf novae (see e.g. \citealt{lasota2001} for a review). The Kepler frequency as a potential counterpart of $f_{\rm b}$ should then decrease. The opposite was observed in MV\,Lyr (\citealt{dobrotka2020}), and in SRGt\,062340.2-265715 in this work. This rules out the inner disc radii as the source.

The central sandwich corona is a result of evaporation (\citealt{meyer1994}). If the mass accretion rate drops during the transition to the low state, also the evaporation of matter to the corona can drop. Such corona can shrink radially with increasing characteristic frequency as a result (top panel of Fig.~3 in \citealt{scaringi2014}). Such radial "shrinking" of the corona was observed in AGNs IRAS\,13224-3809 (\citealt{kara2013}) and 1H\,0707-495 (\citealt{wilkins2014}). The authors conclude that the X-ray source is more compact during low flux states.

Sandwich model is an attractive solution, but the energetic problem due to X-ray to optical luminosity ratio being lower than one (see e.g. \citealt{balman2014,dobrotka2020}) disqualifies the reprocessing scenario in which the optical radiation is generated by reprocessing of X-rays. The sandwich can still exist, but the optical radiation must be generated by the geometrically thin disc itself, and the same $f_{\rm b}$ in optical and X-rays implies common mass accretion fluctuations. Since the central corona is a result of evaporation of the thin disc (\citealt{meyer1994}), any fluctuation in this disc must result in fluctuations in evaporated matter. The common origin is natural then. Another possibility for the corona to play a role in X-ray flickering generation is based on a process known from AGNs. The soft disc photon from a geometrically thin disc irradiates the corona, and the photons are upscattered to observed X-rays by the inverse Compton process (see e.g. \citealt{haardt1991,haardt1993}). If any mass accretion fluctuations characterised by $f_{\rm b}$ are present in the corona, this can leave imprints in the upscattered X-ray radiation.

\section{Summary and conclusions}
\label{section_summary}

We studied optical and X-ray flickering in selected nova-like cataclysmic variables observed by the TESS and \xmm spacecrafts. We searched for $f_{\rm b}$ in the corresponding PDSs.

We found a new optical $f_{\rm b}$ in three systems, and confirmed the previously reported result (\citealt{dobrotka2024}) that the value of $f_{\rm b}$ in nova-likes is clustered around 1\,mHz. The long-term ASAS-SN light curves suggest that the PDS and $f_{\rm b}$ can vary in shape and in value during brightness changes. $f_{\rm b}$ can even disappear. If $f_{\rm b}$ is changing it follows the correlation found by \citet{dobrotka2020}, where $f_{\rm b}$ increases with decreasing brightness. \citet{dobrotka2024} proposed that systems with detected $f_{\rm b}$ have lower inclination than approximately 60-75$^{\circ}$, and that a correlation between WD mass and $f_{\rm b}$ may exist. Thanks to the new measurements in this work, we see no correlation with WD mass, but the inclination condition is confirmed. However, since higher inclination systems tend to be fainter, the non-detection can be a result of limited TESS capabilities. Therefore, the case with systems showing only red noise is not conclusive, but it appears that nova-likes showing $f_{\rm b}$ are not high inclination systems. Exclusive of low brightness, the non-detection of $f_{\rm b}$ for higher inclinations can suggest obscuration of the flickering source being the central disc by the disc edge.

We found two new X-ray counterparts of optical $f_{\rm b}$ in V504\,Cen and V751\,Cyg. The presence of the same $f_{\rm b}$ in optical and X-rays was seen only in MV\,Lyr so far. This implies a similar physical origin of the flickering in more or all nova-likes in general, and points toward a very central disc for source localisation. The X-ray nature of the variability, together with the increasing value of $f_{\rm b}$ with decreasing optical brightness, supports the central sandwich corona as a potential source.

\begin{acknowledgments}
Funded by the EU NextGenerationEU through the Recovery and Resilience Plan for Slovakia under the project No. 09I03-03-V04-00378. This work makes use of ASAS-SN data (\citealt{shappee2014,kochanek2017}). We thank anonymous referee for helpful report and for useful suggestions mainly in the case of inclination vs detection discussion.
\end{acknowledgments}

\begin{contribution}

AD came up with the initial research concept, analysed \xmm\ data and wrote the manuscript. JM provided TESS data analysis and edited the manuscript. MM participated in data handling, figure preparation, and edited the manuscript.


\end{contribution}

%
\facilities{TESS, \xmm}

\software{astropy \citep{astropy_collaboration2013,astropy_collaboration2018,astropy_collaboration2022},
          GNUPLOT (\url{http://www.gnuplot.info/}),
          SAS (\url{https://www.cosmos.esa.int/web/xmm-newton/sas})
          }



\bibliography{mybib}{}

\begin{thebibliography}{}
\expandafter\ifx\csname natexlab\endcsname\relax\def\natexlab#1{#1}\fi
\providecommand{\url}[1]{\href{#1}{#1}}
\providecommand{\dodoi}[1]{doi:~\href{http://doi.org/#1}{\nolinkurl{#1}}}
\providecommand{\doeprint}[1]{\href{http://ascl.net/#1}{\nolinkurl{http://ascl.net/#1}}}
\providecommand{\doarXiv}[1]{\href{https://arxiv.org/abs/#1}{\nolinkurl{https://arxiv.org/abs/#1}}}

\bibitem[{P. {Ar{\'e}valo} \& P. {Uttley}(2006){Ar{\'e}valo} \&
  {Uttley}}]{arevalo2006}
{Ar{\'e}valo}, P., \& {Uttley}, P. 2006, \bibinfo{title}{{Investigating a
  fluctuating-accretion model for the spectral-timing properties of accreting
  black hole systems},} \mnras, 367, 801,
  \dodoi{10.1111/j.1365-2966.2006.09989.x}

\bibitem[{ {Astropy Collaboration} {et~al.}(2013){Astropy Collaboration},
  {Robitaille}, {Tollerud}, {Greenfield}, {Droettboom}, {Bray}, {Aldcroft},
  {Davis}, {Ginsburg}, {Price-Whelan}, {Kerzendorf}, {Conley}, {Crighton},
  {Barbary}, {Muna}, {Ferguson}, {Grollier}, {Parikh}, {Nair}, {Unther},
  {Deil}, {Woillez}, {Conseil}, {Kramer}, {Turner}, {Singer}, {Fox}, {Weaver},
  {Zabalza}, {Edwards}, {Azalee Bostroem}, {Burke}, {Casey}, {Crawford},
  {Dencheva}, {Ely}, {Jenness}, {Labrie}, {Lim}, {Pierfederici}, {Pontzen},
  {Ptak}, {Refsdal}, {Servillat}, \& {Streicher}}]{astropy_collaboration2013}
{Astropy Collaboration}, {Robitaille}, T.~P., {Tollerud}, E.~J., {et~al.} 2013,
  \bibinfo{title}{{Astropy: A community Python package for astronomy},} \aap,
  558, A33, \dodoi{10.1051/0004-6361/201322068}

\bibitem[{ {Astropy Collaboration} {et~al.}(2018){Astropy Collaboration},
  {Price-Whelan}, {Sip{\H{o}}cz}, {G{\"u}nther}, {Lim}, {Crawford}, {Conseil},
  {Shupe}, {Craig}, {Dencheva}, {Ginsburg}, {VanderPlas}, {Bradley},
  {P{\'e}rez-Su{\'a}rez}, {de Val-Borro}, {Aldcroft}, {Cruz}, {Robitaille},
  {Tollerud}, {Ardelean}, {Babej}, {Bach}, {Bachetti}, {Bakanov}, {Bamford},
  {Barentsen}, {Barmby}, {Baumbach}, {Berry}, {Biscani}, {Boquien}, {Bostroem},
  {Bouma}, {Brammer}, {Bray}, {Breytenbach}, {Buddelmeijer}, {Burke},
  {Calderone}, {Cano Rodr{\'\i}guez}, {Cara}, {Cardoso}, {Cheedella}, {Copin},
  {Corrales}, {Crichton}, {D'Avella}, {Deil}, {Depagne}, {Dietrich}, {Donath},
  {Droettboom}, {Earl}, {Erben}, {Fabbro}, {Ferreira}, {Finethy}, {Fox},
  {Garrison}, {Gibbons}, {Goldstein}, {Gommers}, {Greco}, {Greenfield},
  {Groener}, {Grollier}, {Hagen}, {Hirst}, {Homeier}, {Horton}, {Hosseinzadeh},
  {Hu}, {Hunkeler}, {Ivezi{\'c}}, {Jain}, {Jenness}, {Kanarek}, {Kendrew},
  {Kern}, {Kerzendorf}, {Khvalko}, {King}, {Kirkby}, {Kulkarni}, {Kumar},
  {Lee}, {Lenz}, {Littlefair}, {Ma}, {Macleod}, {Mastropietro}, {McCully},
  {Montagnac}, {Morris}, {Mueller}, {Mumford}, {Muna}, {Murphy}, {Nelson},
  {Nguyen}, {Ninan}, {N{\"o}the}, {Ogaz}, {Oh}, {Parejko}, {Parley}, {Pascual},
  {Patil}, {Patil}, {Plunkett}, {Prochaska}, {Rastogi}, {Reddy Janga},
  {Sabater}, {Sakurikar}, {Seifert}, {Sherbert}, {Sherwood-Taylor}, {Shih},
  {Sick}, {Silbiger}, {Singanamalla}, {Singer}, {Sladen}, {Sooley},
  {Sornarajah}, {Streicher}, {Teuben}, {Thomas}, {Tremblay}, {Turner},
  {Terr{\'o}n}, {van Kerkwijk}, {de la Vega}, {Watkins}, {Weaver}, {Whitmore},
  {Woillez}, {Zabalza}, \& {Astropy Contributors}}]{astropy_collaboration2018}
{Astropy Collaboration}, {Price-Whelan}, A.~M., {Sip{\H{o}}cz}, B.~M., {et~al.}
  2018, \bibinfo{title}{{The Astropy Project: Building an Open-science Project
  and Status of the v2.0 Core Package},} \aj, 156, 123,
  \dodoi{10.3847/1538-3881/aabc4f}

\bibitem[{ {Astropy Collaboration} {et~al.}(2022){Astropy Collaboration},
  {Price-Whelan}, {Lim}, {Earl}, {Starkman}, {Bradley}, {Shupe}, {Patil},
  {Corrales}, {Brasseur}, {N{\"o}the}, {Donath}, {Tollerud}, {Morris},
  {Ginsburg}, {Vaher}, {Weaver}, {Tocknell}, {Jamieson}, {van Kerkwijk},
  {Robitaille}, {Merry}, {Bachetti}, {G{\"u}nther}, {Aldcroft},
  {Alvarado-Montes}, {Archibald}, {B{\'o}di}, {Bapat}, {Barentsen},
  {Baz{\'a}n}, {Biswas}, {Boquien}, {Burke}, {Cara}, {Cara}, {Conroy},
  {Conseil}, {Craig}, {Cross}, {Cruz}, {D'Eugenio}, {Dencheva}, {Devillepoix},
  {Dietrich}, {Eigenbrot}, {Erben}, {Ferreira}, {Foreman-Mackey}, {Fox},
  {Freij}, {Garg}, {Geda}, {Glattly}, {Gondhalekar}, {Gordon}, {Grant},
  {Greenfield}, {Groener}, {Guest}, {Gurovich}, {Handberg}, {Hart},
  {Hatfield-Dodds}, {Homeier}, {Hosseinzadeh}, {Jenness}, {Jones}, {Joseph},
  {Kalmbach}, {Karamehmetoglu}, {Ka{\l}uszy{\'n}ski}, {Kelley}, {Kern},
  {Kerzendorf}, {Koch}, {Kulumani}, {Lee}, {Ly}, {Ma}, {MacBride}, {Maljaars},
  {Muna}, {Murphy}, {Norman}, {O'Steen}, {Oman}, {Pacifici}, {Pascual},
  {Pascual-Granado}, {Patil}, {Perren}, {Pickering}, {Rastogi}, {Roulston},
  {Ryan}, {Rykoff}, {Sabater}, {Sakurikar}, {Salgado}, {Sanghi}, {Saunders},
  {Savchenko}, {Schwardt}, {Seifert-Eckert}, {Shih}, {Jain}, {Shukla}, {Sick},
  {Simpson}, {Singanamalla}, {Singer}, {Singhal}, {Sinha}, {Sip{\H{o}}cz},
  {Spitler}, {Stansby}, {Streicher}, {{\v{S}}umak}, {Swinbank}, {Taranu},
  {Tewary}, {Tremblay}, {Val-Borro}, {Van Kooten}, {Vasovi{\'c}}, {Verma}, {de
  Miranda Cardoso}, {Williams}, {Wilson}, {Winkel}, {Wood-Vasey}, {Xue},
  {Yoachim}, {Zhang}, {Zonca}, \& {Astropy Project
  Contributors}}]{astropy_collaboration2022}
{Astropy Collaboration}, {Price-Whelan}, A.~M., {Lim}, P.~L., {et~al.} 2022,
  \bibinfo{title}{{The Astropy Project: Sustaining and Growing a
  Community-oriented Open-source Project and the Latest Major Release (v5.0) of
  the Core Package},} \apj, 935, 167, \dodoi{10.3847/1538-4357/ac7c74}

\bibitem[{A. {Aungwerojwit} {et~al.}(2005){Aungwerojwit}, {G{\"a}nsicke},
  {Rodr{\'\i}guez-Gil}, {Hagen}, {Harlaftis}, {Papadimitriou}, {Lehto},
  {Araujo-Betancor}, {Heber}, {Fried}, {Engels}, \&
  {Katajainen}}]{aungwerojwit2005}
{Aungwerojwit}, A., {G{\"a}nsicke}, B.~T., {Rodr{\'\i}guez-Gil}, P., {et~al.}
  2005, \bibinfo{title}{{HS 0139+0559, HS 0229+8016, HS 0506+7725, and HS
  0642+5049: four new long-period cataclysmic variables},} \aap, 443, 995,
  \dodoi{10.1051/0004-6361:20042610}

\bibitem[{{\c S}. {Balman} {et~al.}(2014){Balman}, {Godon}, \&
  {Sion}}]{balman2014}
{Balman}, {\c S}., {Godon}, P., \& {Sion}, E.~M. 2014, \bibinfo{title}{{Swift
  X-Ray Telescope Observations of the Nova-like Cataclysmic Variables MV Lyr,
  BZ Cam, and V592 Cas},} \apj, 794, 84, \dodoi{10.1088/0004-637X/794/1/84}

\bibitem[{{\c S}. {Balman} \& M. {Revnivtsev}(2012){Balman} \&
  {Revnivtsev}}]{balman2012}
{Balman}, {\c S}., \& {Revnivtsev}, M. 2012, \bibinfo{title}{{X-ray variations
  in the inner accretion flow of dwarf novae},} \aap, 546, A112,
  \dodoi{10.1051/0004-6361/201219469}

\bibitem[{A. {Bruch}(2022){Bruch}}]{bruch2022}
{Bruch}, A. 2022, \bibinfo{title}{{TESS light curves of cataclysmic variables -
  I - Unknown periods in long-known stars},} \mnras, 514, 4718,
  \dodoi{10.1093/mnras/stac1650}

\bibitem[{A. {Bruch}(2023{\natexlab{a}}){Bruch}}]{bruch2023a}
{Bruch}, A. 2023{\natexlab{a}}, \bibinfo{title}{{TESS light curves of
  cataclysmic variables - II - Superhumps in old novae and novalike
  variables},} \mnras, 519, 352, \dodoi{10.1093/mnras/stac3493}

\bibitem[{A. {Bruch}(2023{\natexlab{b}}){Bruch}}]{bruch2023b}
{Bruch}, A. 2023{\natexlab{b}}, \bibinfo{title}{{TESS light curves of
  cataclysmic variables - III - More superhump systems among old novae and
  nova-like variables},} \mnras, 525, 1953, \dodoi{10.1093/mnras/stad2089}

\bibitem[{D.~A.~H. {Buckley} \& I.~R. {Tuohy}(1990){Buckley} \&
  {Tuohy}}]{buckley1990}
{Buckley}, D.~A.~H., \& {Tuohy}, I.~R. 1990, \bibinfo{title}{{H0534-581: A New
  Intermediate Polar?},} \apj, 349, 296, \dodoi{10.1086/168314}

\bibitem[{A. {Dobrotka} {et~al.}(2024){Dobrotka}, {Magdolen}, \&
  {Jan{\'\i}kov{\'a}}}]{dobrotka2024}
{Dobrotka}, A., {Magdolen}, J., \& {Jan{\'\i}kov{\'a}}, D. 2024,
  \bibinfo{title}{{Searching for the mHz variability in the TESS observations
  of nova-like cataclysmic variables},} \aap, 692, A27,
  \dodoi{10.1051/0004-6361/202451004}

\bibitem[{A. {Dobrotka} {et~al.}(2014){Dobrotka}, {Mineshige}, \&
  {Ness}}]{dobrotka2014}
{Dobrotka}, A., {Mineshige}, S., \& {Ness}, J.-U. 2014,
  \bibinfo{title}{{Resolving different sources of fast X-ray variability of the
  dwarf nova RU Peg in quiescence},} \mnras, 438, 1714,
  \dodoi{10.1093/mnras/stt2311}

\bibitem[{A. {Dobrotka} {et~al.}(2020){Dobrotka}, {Negoro}, \&
  {Konopka}}]{dobrotka2020}
{Dobrotka}, A., {Negoro}, H., \& {Konopka}, P. 2020,
  \bibinfo{title}{{Alternation of the flickering morphology between the high
  and low state in MV Lyrae},} \aap, 641, A55,
  \dodoi{10.1051/0004-6361/201935569}

\bibitem[{A. {Dobrotka} {et~al.}(2019){Dobrotka}, {Negoro}, \&
  {Mineshige}}]{dobrotka2019}
{Dobrotka}, A., {Negoro}, H., \& {Mineshige}, S. 2019, \bibinfo{title}{{Similar
  shot profile morphology of fast variability in a cataclysmic variable, X-ray
  binary, and blazar: The MV Lyrae case},} \aap, 631, A134,
  \dodoi{10.1051/0004-6361/201935198}

\bibitem[{A. {Dobrotka} \& J.-U. {Ness}(2015){Dobrotka} \&
  {Ness}}]{dobrotka2015}
{Dobrotka}, A., \& {Ness}, J.-U. 2015, \bibinfo{title}{{Differences in the fast
  optical variability of the dwarf nova V1504 Cyg between quiescence and
  outbursts detected in Kepler data and simulations of the rms-flux
  relations},} \mnras, 451, 2851, \dodoi{10.1093/mnras/stv1178}

\bibitem[{A. {Dobrotka} {et~al.}(2016){Dobrotka}, {Ness}, \& {Baj{\v c}i{\v
  c}{\'a}kov{\'a}}}]{dobrotka2016}
{Dobrotka}, A., {Ness}, J.-U., \& {Baj{\v c}i{\v c}{\'a}kov{\'a}}, I. 2016,
  \bibinfo{title}{{Fast stochastic variability study of two SU UMa systems
  V1504 Cyg and V344 Lyr observed by Kepler satellite},} \mnras, 460, 458,
  \dodoi{10.1093/mnras/stw1001}

\bibitem[{A. {Dobrotka} {et~al.}(2017){Dobrotka}, {Ness}, {Mineshige}, \&
  {Nucita}}]{dobrotka2017}
{Dobrotka}, A., {Ness}, J.-U., {Mineshige}, S., \& {Nucita}, A.~A. 2017,
  \bibinfo{title}{{XMM-Newton observation of MV Lyr and the sandwiched model
  confirmation},} \mnras, 468, 1183, \dodoi{10.1093/mnras/stx513}

\bibitem[{A. {Dobrotka} {et~al.}(2021){Dobrotka}, {Orio}, {Benka}, \&
  {Vanderburg}}]{dobrotka2021}
{Dobrotka}, A., {Orio}, M., {Benka}, D., \& {Vanderburg}, A. 2021,
  \bibinfo{title}{{Searching for the 1 mHz variability in the flickering of
  V4743 Sgr: A cataclysmic variable accreting at a high rate},} \aap, 649, A67,
  \dodoi{10.1051/0004-6361/202039742}

\bibitem[{J.~B. {Dove} {et~al.}(1997){Dove}, {Wilms}, {Maisack}, \&
  {Begelman}}]{dove1997}
{Dove}, J.~B., {Wilms}, J., {Maisack}, M., \& {Begelman}, M.~C. 1997,
  \bibinfo{title}{{Self-consistent Thermal Accretion Disk Corona Models for
  Compact Objects. II. Application to Cygnus X-1},} \apj, 487, 759,
  \dodoi{10.1086/304647}

\bibitem[{P. {Godon} {et~al.}(2007){Godon}, {Sion}, {Barrett}, \&
  {Szkody}}]{godon2007}
{Godon}, P., {Sion}, E.~M., {Barrett}, P., \& {Szkody}, P. 2007,
  \bibinfo{title}{{A Far-Ultraviolet Study of the Nova-like V794 Aquilae},}
  \apj, 656, 1092, \dodoi{10.1086/510775}

\bibitem[{R. {H{\= o}shi}(1979){H{\= o}shi}}]{hoshi1979}
{H{\= o}shi}, R. 1979, \bibinfo{title}{{Accretion Model for Outbursts of Dwarf
  Nova},} Progress of Theoretical Physics, 61, 1307,
  \dodoi{10.1143/PTP.61.1307}

\bibitem[{F. {Haardt} \& L. {Maraschi}(1991){Haardt} \&
  {Maraschi}}]{haardt1991}
{Haardt}, F., \& {Maraschi}, L. 1991, \bibinfo{title}{{A Two-Phase Model for
  the X-Ray Emission from Seyfert Galaxies},} \apjl, 380, L51,
  \dodoi{10.1086/186171}

\bibitem[{F. {Haardt} \& L. {Maraschi}(1993){Haardt} \&
  {Maraschi}}]{haardt1993}
{Haardt}, F., \& {Maraschi}, L. 1993, \bibinfo{title}{{X-Ray Spectra from
  Two-Phase Accretion Disks},} \apj, 413, 507, \dodoi{10.1086/173020}

\bibitem[{R.~K. {Honeycutt}(2001){Honeycutt}}]{honeycutt2001}
{Honeycutt}, R.~K. 2001, \bibinfo{title}{{Similarities between Stunted
  Outbursts in Nova-like Cataclysmic Variables and Outbursts in Ordinary Dwarf
  Novae},} \pasp, 113, 473, \dodoi{10.1086/319543}

\bibitem[{R.~K. {Honeycutt} {et~al.}(1994){Honeycutt}, {Cannizzo}, \&
  {Robertson}}]{honeycutt1994}
{Honeycutt}, R.~K., {Cannizzo}, J.~K., \& {Robertson}, J.~W. 1994,
  \bibinfo{title}{{The High-State/Low-State Transition in V794 Aquilae},} \apj,
  425, 835, \dodoi{10.1086/174028}

\bibitem[{R.~K. {Honeycutt} \& S. {Kafka}(2004){Honeycutt} \&
  {Kafka}}]{honeycutt2004}
{Honeycutt}, R.~K., \& {Kafka}, S. 2004, \bibinfo{title}{{Characteristics of
  High-State/Low-State Transitions in VY Sculptoris Stars},} \aj, 128, 1279,
  \dodoi{10.1086/422737}

\bibitem[{R.~K. {Honeycutt} \& J.~W. {Robertson}(1998){Honeycutt} \&
  {Robertson}}]{honeycutt1998}
{Honeycutt}, R.~K., \& {Robertson}, J.~W. 1998, \bibinfo{title}{{Multiyear
  Photometry and a Spectroscopic Orbital Period Search for the VY SCULPTORIS
  Type Cataclysmic Variable V794 Aquilae},} \aj, 116, 1961,
  \dodoi{10.1086/300539}

\bibitem[{E. {Kara} {et~al.}(2013){Kara}, {Fabian}, {Cackett}, {Miniutti}, \&
  {Uttley}}]{kara2013}
{Kara}, E., {Fabian}, A.~C., {Cackett}, E.~M., {Miniutti}, G., \& {Uttley}, P.
  2013, \bibinfo{title}{{Revealing the X-ray source in IRAS 13224-3809 through
  flux-dependent reverberation lags},} \mnras, 430, 1408,
  \dodoi{10.1093/mnras/stt024}

\bibitem[{C.~S. {Kochanek} {et~al.}(2017){Kochanek}, {Shappee}, {Stanek},
  {Holoien}, {Thompson}, {Prieto}, {Dong}, {Shields}, {Will}, {Britt},
  {Perzanowski}, \& {Pojma{\'n}ski}}]{kochanek2017}
{Kochanek}, C.~S., {Shappee}, B.~J., {Stanek}, K.~Z., {et~al.} 2017,
  \bibinfo{title}{{The All-Sky Automated Survey for Supernovae (ASAS-SN) Light
  Curve Server v1.0},} \pasp, 129, 104502, \dodoi{10.1088/1538-3873/aa80d9}

\bibitem[{O. {Kotov} {et~al.}(2001){Kotov}, {Churazov}, \&
  {Gilfanov}}]{kotov2001}
{Kotov}, O., {Churazov}, E., \& {Gilfanov}, M. 2001, \bibinfo{title}{{On the
  X-ray time-lags in the black hole candidates},} \mnras, 327, 799,
  \dodoi{10.1046/j.1365-8711.2001.04769.x}

\bibitem[{J.-P. {Lasota}(2001){Lasota}}]{lasota2001}
{Lasota}, J.-P. 2001, \bibinfo{title}{{The disc instability model of dwarf
  novae and low-mass X-ray binary transients},} \nar, 45, 449,
  \dodoi{10.1016/S1387-6473(01)00112-9}

\bibitem[{A.~P. {Linnell} {et~al.}(2009){Linnell}, {Godon}, {Hubeny}, {Sion},
  {Szkody}, \& {Barrett}}]{linnell2009}
{Linnell}, A.~P., {Godon}, P., {Hubeny}, I., {et~al.} 2009,
  \bibinfo{title}{{V3885 Sagittarius: A Comparison with a Range of Standard
  Model Accretion Disks},} \apj, 703, 1839,
  \dodoi{10.1088/0004-637X/703/2/1839}

\bibitem[{Y.~E. {Lyubarskii}(1997){Lyubarskii}}]{lyubarskii1997}
{Lyubarskii}, Y.~E. 1997, \bibinfo{title}{{Flicker noise in accretion discs},}
  \mnras, 292, 679

\bibitem[{F. {Meyer} \& E. {Meyer-Hofmeister}(1981){Meyer} \&
  {Meyer-Hofmeister}}]{meyer1981}
{Meyer}, F., \& {Meyer-Hofmeister}, E. 1981, \bibinfo{title}{{On the Elusive
  Cause of Cataclysmic Variable Outbursts},} \aap, 104, L10

\bibitem[{F. {Meyer} \& E. {Meyer-Hofmeister}(1994){Meyer} \&
  {Meyer-Hofmeister}}]{meyer1994}
{Meyer}, F., \& {Meyer-Hofmeister}, E. 1994, \bibinfo{title}{{Accretion disk
  evaporation by a coronal siphon flow.},} \aap, 288, 175

\bibitem[{Y. {Osaki}(1974){Osaki}}]{osaki1974}
{Osaki}, Y. 1974, \bibinfo{title}{{An accretion model for the outbursts of U
  Geminorum stars},} \pasj, 26, 429

\bibitem[{I.~E. {Papadakis} \& A. {Lawrence}(1993){Papadakis} \&
  {Lawrence}}]{papadakis1993}
{Papadakis}, I.~E., \& {Lawrence}, A. 1993, \bibinfo{title}{{Improved Methods
  for Power Spectrum Modelling of Red Noise},} \mnras, 261, 612,
  \dodoi{10.1093/mnras/261.3.612}

\bibitem[{F.~M.~A. {Ribeiro} \& M.~P. {Diaz}(2007){Ribeiro} \&
  {Diaz}}]{ribeiro2007}
{Ribeiro}, F. M.~A., \& {Diaz}, M.~P. 2007, \bibinfo{title}{{Emission-Line
  Flickering from the Secondary Star in Cataclysmic Variables? A Study of V3885
  Sagittarii},} \aj, 133, 2659, \dodoi{10.1086/514335}

\bibitem[{H. {Ritter} \& U. {Kolb}(2003){Ritter} \& {Kolb}}]{ritter2003}
{Ritter}, H., \& {Kolb}, U. 2003, \bibinfo{title}{{Catalogue of cataclysmic
  binaries, low-mass X-ray binaries and related objects (Seventh edition)},}
  \aap, 404, 301, \dodoi{10.1051/0004-6361:20030330}

\bibitem[{P. {Rodr{\'\i}guez-Gil} {et~al.}(2007){Rodr{\'\i}guez-Gil},
  {Schmidtobreick}, \& {G{\"a}nsicke}}]{rodriguez2007}
{Rodr{\'\i}guez-Gil}, P., {Schmidtobreick}, L., \& {G{\"a}nsicke}, B.~T. 2007,
  \bibinfo{title}{{Spectroscopic search for new SW Sextantis stars in the 3-4 h
  orbital period range - I},} \mnras, 374, 1359,
  \dodoi{10.1111/j.1365-2966.2006.11245.x}

\bibitem[{J.~D. {Scargle}(1982){Scargle}}]{scargle1982}
{Scargle}, J.~D. 1982, \bibinfo{title}{{Studies in astronomical time series
  analysis. II - Statistical aspects of spectral analysis of unevenly spaced
  data},} \apj, 263, 835, \dodoi{10.1086/160554}

\bibitem[{S. {Scaringi}(2014){Scaringi}}]{scaringi2014}
{Scaringi}, S. 2014, \bibinfo{title}{{A physical model for the flickering
  variability in cataclysmic variables},} \mnras, 438, 1233,
  \dodoi{10.1093/mnras/stt2270}

\bibitem[{S. {Scaringi} {et~al.}(2012){Scaringi}, {K{\"o}rding}, {Uttley},
  {Groot}, {Knigge}, {Still}, \& {Jonker}}]{scaringi2012a}
{Scaringi}, S., {K{\"o}rding}, E., {Uttley}, P., {et~al.} 2012,
  \bibinfo{title}{{Broad-band timing properties of the accreting white dwarf MV
  Lyrae},} \mnras, 427, 3396, \dodoi{10.1111/j.1365-2966.2012.22022.x}

\bibitem[{M.~R. {Schreiber} {et~al.}(2003){Schreiber}, {Hameury}, \&
  {Lasota}}]{schreiber2003}
{Schreiber}, M.~R., {Hameury}, J.~M., \& {Lasota}, J.~P. 2003,
  \bibinfo{title}{{Delays in dwarf novae I: The case of SS Cygni},} \aap, 410,
  239, \dodoi{10.1051/0004-6361:20031221}

\bibitem[{B. {Shappee} {et~al.}(2014){Shappee}, {Prieto}, {Stanek}, {Kochanek},
  {Holoien}, {Jencson}, {Basu}, {Beacom}, {Szczygiel}, {Pojmanski},
  {Brimacombe}, {Dubberley}, {Elphick}, {Foale}, {Hawkins}, {Mullins},
  {Rosing}, {Ross}, \& {Walker}}]{shappee2014}
{Shappee}, B., {Prieto}, J., {Stanek}, K.~Z., {et~al.} 2014,
  \bibinfo{title}{{All Sky Automated Survey for SuperNovae (ASAS-SN or
  ``Assassin'')},} in American Astronomical Society Meeting Abstracts, Vol.
  223, American Astronomical Society Meeting Abstracts \#223, 236.03

\bibitem[{B. {Warner}(1995){Warner}}]{warner1995}
{Warner}, B. 1995, \bibinfo{title}{{Cataclysmic variable stars.},} Cambridge
  Astrophysics Series, 28

\bibitem[{D.~R. {Wilkins} {et~al.}(2014){Wilkins}, {Kara}, {Fabian}, \&
  {Gallo}}]{wilkins2014}
{Wilkins}, D.~R., {Kara}, E., {Fabian}, A.~C., \& {Gallo}, L.~C. 2014,
  \bibinfo{title}{{Caught in the act: measuring the changes in the corona that
  cause the extreme variability of 1H 0707-495},} \mnras, 443, 2746,
  \dodoi{10.1093/mnras/stu1273}

\end{thebibliography}
\bibliographystyle{aasjournalv7}



\end{document}